\documentclass[conference]{IEEEtran}
\IEEEoverridecommandlockouts
\usepackage{cite}
\usepackage{url}
\usepackage{booktabs}
\usepackage{amsmath,amssymb,amsfonts}
\usepackage{algorithmic}
\usepackage{graphicx}
\usepackage{lipsum}
\usepackage[T1]{fontenc}
\usepackage{textcomp}
\usepackage{tikz}
\usepackage{atbegshi}
\usepackage{xcolor}
\usepackage{csquotes}
\usepackage{comment}
\usepackage{hyperref}
 \hypersetup{
colorlinks=true,
linkcolor=blue,
filecolor=blue,
citecolor = blue,
urlcolor=blue,
}
\def\BibTeX{{\rm B\kern-.05em{\sc i\kern-.025em b}\kern-.08em
    T\kern-.1667em\lower.7ex\hbox{E}\kern-.125emX}}
\begin{document}

\title{Optimized IoT Intrusion Detection using Machine Learning Technique}

\author{
\IEEEauthorblockN{Muhammad Zawad Mahmud}
\IEEEauthorblockA{\textit{Dept. of ECE} \\
\textit{North South University}\\
Dhaka-1229, Bangladesh \\
zawad.mahmud1@northsouth.edu}
\and
\IEEEauthorblockN{Samiha Islam}
\IEEEauthorblockA{\textit{Dept. of ECE} \\ 
\textit{North South University}\\
Dhaka-1229, Bangladesh \\
samiha.islam2@northsouth.edu}
\and
\IEEEauthorblockN{Shahran Rahman Alve}
\IEEEauthorblockA{\textit{Dept. of ECE} \\
\textit{North South University}\\
Dhaka-1229, Bangladesh \\
shahran.alve@northsouth.edu}
\and 
\IEEEauthorblockN{Al Jubayer Pial}
\IEEEauthorblockA{\textit{Dept. of ECE} \\
\textit{North South University}\\
Dhaka-1229, Bangladesh \\
jubayer.pial@northsouth.edu}
}

\maketitle

\begin{abstract}
An application of software known as an Intrusion Detection System (IDS) employs machine algorithms to identify network intrusions. Selective logging, safeguarding privacy, reputation-based defense against numerous attacks, and dynamic response to threats are a few of the problems that intrusion identification is used to solve. The biological system known as IoT has seen a rapid increase in high dimensionality and information traffic. Self-protective mechanisms like intrusion detection systems (IDSs) are essential for defending against a variety of attacks. On the other hand, the functional and physical diversity of IoT IDS systems causes significant issues. These attributes make it troublesome and unrealistic to completely use all IoT elements and properties for IDS self-security. For peculiarity-based IDS, this study proposes and implements a novel component selection and extraction strategy (our strategy). A five-ML algorithm model-based IDS for machine learning-based networks with proper hyperparamater tuning is presented in this paper by examining how the most popular feature selection methods and classifiers are combined, such as K-Nearest Neighbors (KNN) Classifier, Decision Tree (DT) Classifier, Random Forest (RF) Classifier, Gradient Boosting Classifier, and Ada Boost Classifier. The Random Forest (RF) classifier had the highest accuracy of 99.39\%. The K-Nearest Neighbor (KNN) classifier exhibited the lowest performance among the evaluated models, achieving an accuracy of 94.84\%. This study's models have a significantly higher performance rate than those used in previous studies, indicating that they are more reliable.
\end{abstract}


\section{Introduction}
Attacks on computer networks and systems, as well as on information systems, are the goal of intrusion detection systems. Indeed, it is challenging to establish information systems that can be shown to be secure and to safeguard them throughout their existence and use. A completely secure information system cannot always be created due to operational or historical constraints. Digital technology is being used more and more in all areas of life, especially business. The seamless collection and sharing of data amongst networked devices has been made possible by the Internet of Things (IoT), revolutionizing a number of sectors. Significant advantages of this connectedness include increased productivity, automation, and data-driven decision-making~\cite{khan2012future}. The proliferation of IoT devices does, however, also present serious security risks because these devices are usually vulnerable to various cyber threats and assaults. As to a Kaspersky analysis, there was a 100\% surge in IoT assaults during the first half of 2021, with 1.51 billion breaches, as opposed to 639 million during the same time in 2020~\cite{kaspersky2021}.

Traditional intrusion detection techniques have drawbacks, including high false-positive rates, scaling problems, and difficulty adjusting to novel and emerging threats, despite attempts to safeguard IoT networks. Thus, the need for sophisticated safety measures that can successfully shield IoT networks from malicious activity is urgent.

By using machine learning techniques to construct an intrusion detection system (IDS), this research seeks to overcome these issues. A potent method for sifting through massive amounts of data, finding patterns, and spotting abnormalities that can point to security breaches is machine learning. Our objective is to develop IDS for IoT settings by utilizing cutting-edge machine learning-based methods. The primary accomplishments of this work are as follows:
\begin{itemize}
\item Utilizing the machine learning methods with hyperparameter tuning with the help of, the research aims to extract essential features from a dataset containing both attacks and normal situations.
\item With a testing accuracy of 99.39\%, the proposed Random Forest with hyperparamter tuning aims to create a dependable, economical, and rapid diagnostic solution characterized by high accuracy and robust validation measures.
\end{itemize}
The novelty of this work lies in the use of Random Forest with proper tuning and the testing accuracy of 99.39\% better than the existing ones available.

\section{Related Work}
The Industrial Internet of Things (IIoT) has revolutionized industrial processes by using sensor data to optimize operations and increase efficiency. Nevertheless, the expansion of IIoT has resulted in significant cybersecurity challenges, particularly sophisticated cyber-attacks that threaten the integrity, confidentiality, and availability of critical industrial systems. In order to find possible security breaches, IoT (Internet of Things) intrusion detection utilizing machine learning and deep learning models for numeric datasets analyzes patterns and abnormalities in the data gathered from IoT devices. Researchers are attracted to these types of solutions because these algorithms can handle gigantic data over less time and can identify even harder-to-track intrusion categories impossible for older methods. Since these models are self-learning by nature of machine learning, they can add even further robustness to IoT environments.

Potnurwar et al.~\cite{potnurwar2023deep}, with an aim to tackle these problems, present a novel deep-learning-based intrusion detection method and hybrid feature selection, among other methods motivated by the deep forward neural network (DFFNN) and a rule-based hybrid feature selection. The research uses the NSL-KDD and UNSW-NB15 datasets. With its best-performing test using F1 scores, researchers derived a true positive rate of very high (96\% in attack and 98\% in normal).

Mohy-Eddine et al.~\cite{mohy2023ensemble} presented a new intrusion detection model towards a more secured Industrial Internet of Things (IIoT); it analyzed several of the vulnerabilities that are typical in the IIoT environments. The method utilizes Isolation Forest (IF) and Pearson Correlation Coefficient (PCC) in the feature engineering and applies Random Forest (RF) classifiers to enhance the detection performance. The results presented above highlight that the performance of the RF-PCCIF and RF-IFPCC configurations was excellent when using the NF-UNSW-NB15-v2 datasets, with RF-PCCIF achieving 99.30\%.

A novel intrusion detection model for IoT networks that can resolve the shortcomings of the traditional firewalls and encryption techniques to cope with the increasing network threats was introduced by Zakariah et al.~\cite{zakariah2023machine}. They have done research based on convolutional neural networks (CNN) and long short-term memory (LSTM) networks, castled with attention and complemented with adaptive synthetic sampling (ADASYN) to tackle imbalanced datasets. The results demonstrate that the hybrid model used performed at AUC = 0.98, accuracy = 0.89, precision = 0.90, recall = 0.89, and f1 = 0.90, outperforming the MLP baseline, which indicates AUC = 0.88, accuracy = 0.87, precision = 0.89, recall = 0.89, and f1 = 0.88.

Chaganti et al.~\cite{chaganti2023deep} deliver an improved intrusion detection system based on deep learning with long short-term memory (LSTM) networks to protect SDNL-enabled IoT networks, which strengthen the detection and classification of attacks. The proposed LSTM architecture could classify each multi-class attack with 97.7\% accuracy after applying it to two datasets involving SDNIoT and different network threat techniques (the LSTM model can avoid DDoS, surveillance, and other network attacks).

Nowadays, many IoT devices are developed based on low-power and lossy networks, like the Routing Protocol for Low-Power and Lossy Networks (RPL); this led to a great opportunity to launch many different kinds of security attacks as these devices have limited computational resources. Because of the inherent constraints of these networks, the traditional IDS method failed to effectively monitor and secure the traffic of IoT devices. Al Sawafi et al.~\cite{al2023hybrid} addressed the matter and proposed deep learning-based IDS mechanisms that combined both supervised and semi-supervised classification algorithms to separate the group of network attacks that are based on the type of known and type of unknown attacks. Results show that their deep learning-based model secured a 99\% accuracy rate and 98\% F1-score in detecting known attacks, highlighting the contribution of deep learning to enhance IoT security.

Rose et al.~\cite{rose2021intrusion} in their research titled 'Intrusion Detection using Network Traffic Profiling and Machine Learning for IoT' generated a dataset and model to test the effectiveness of network characterization and machine learning in protecting IoT devices from cyber-attacks. The experimental results show that the proposed anomaly detection system achieves 98.35\% accuracy and 98.35\% false-positive alarms.

Musleh et al.~\cite{musleh2023intrusion} emphasize the importance of feature extraction in improving the efficacy of ML-based intrusion detection systems (IDS) for IoT networks. Using the IEEE Dataport dataset, the paper demonstrates that the VGG-16 model, when paired with a stacked ensemble classifier including KNN and sequential minimum optimization (SMO), achieves an impressive 98.3\% accuracy while preserving ideal precision.

Traditional intrusion detection systems (IDS) frequently focus on a single layer of the IoT architecture, limiting their overall efficacy. To address these shortcomings, Khan et al.~\cite{khan2023hybrid} present a hybrid deep learning-based intrusion detection system that uses recurrent neural networks and gated recurrent units to identify intrusions across IoT systems' physical, network, and application levels. The proposed model, which was evaluated on the ToN-IoT dataset, includes advanced deep learning algorithms optimized with the Adam and Adamax approaches. Results of the evaluation show high accuracy, precision, recall, and F1-score equal to 0.97, 0.96, 0.95, and 0.96, respectively, when the optimizer is Adamax.

In this study, we used five different machine learning classifier models on the BoTNeTIoT-L01 dataset. We used those machine learning classifier models in our dataset after splitting the data into an 80-20 ratio. 80\% of the data was used for training, and the remaining 20\% was used for test data, which was kept separately. Among the models, the Random Forest classifier gave the highest accuracy of 99.39\% and an F1 score of 99.4\%. As discussed above, no other study has been able to reach this level of accuracy for IoT intrusion detection.

\section{Methodology}
\subsection{Dataset}
\begin{figure}[h]
    \centering
    \includegraphics[width=0.45\textwidth] {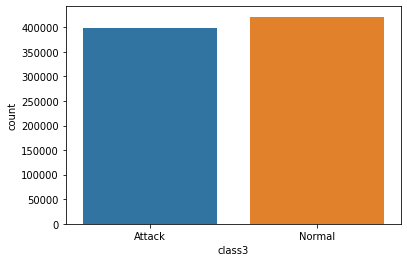}
    \caption{Dataset}
    \label{fig:1}
\end{figure}

An Internet of Things intrusion detection system dataset~\cite{azalhowaide_iot_2023} provided the basis for the research. There are 820834 rows and 54 columns in this dataset. The value of the output column, class 3, is either one or zero. If the value is 0, an attack has not been detected, and if it is 1, an assault has been detected. As depicted in Fig.~\ref{fig:1}, the data in this set are balanced. The data was split into 80-20 ratios for training and testing, respectively.
\subsection{Model Description}
\begin{figure}[h]
    \centering
    \includegraphics[width=0.45\textwidth]{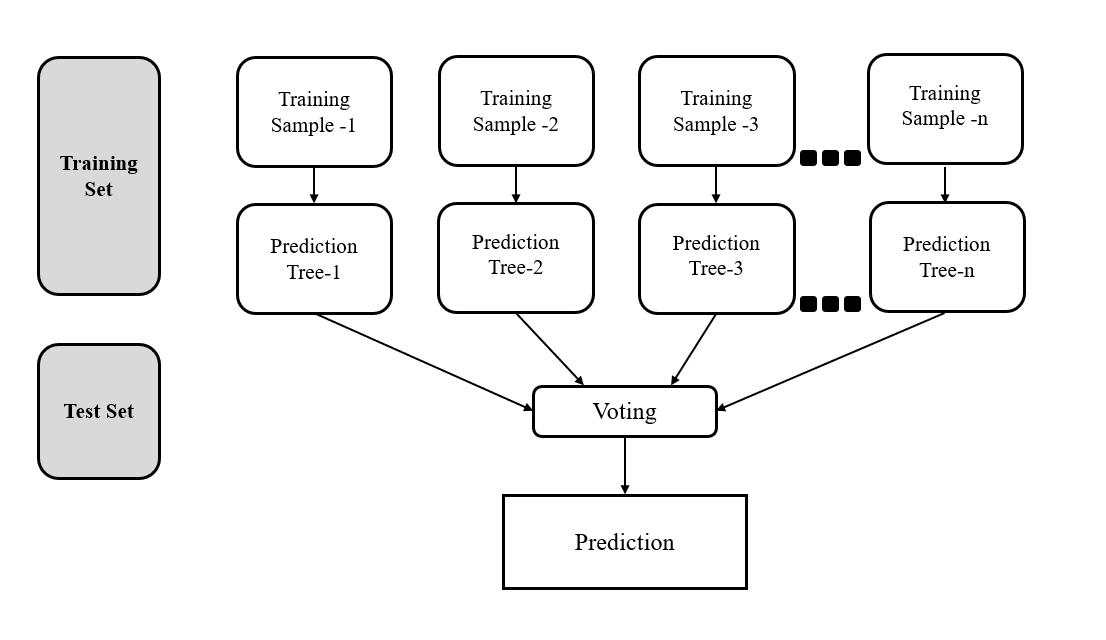}
    \caption{Diagram of Random Forest Classifier}
    \label{fig:3}
\end{figure}

\begin{figure}[h]
    \centering
    \includegraphics[width=0.45\textwidth]{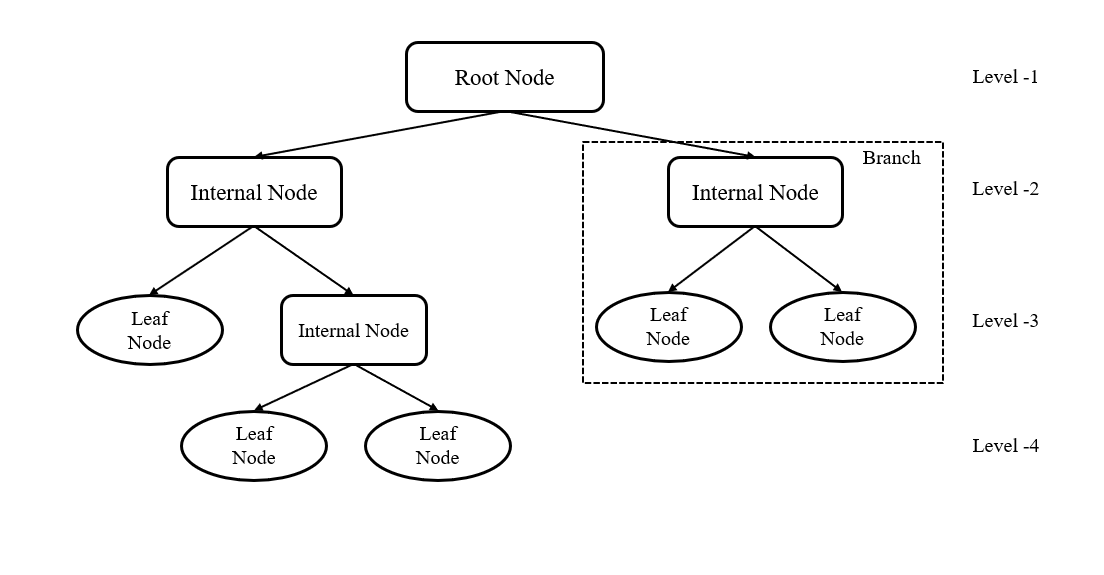}
    \caption{Diagram of Decision Tree Classifier}
    \label{fig:4}
\end{figure}

The first algorithm used was the Random Forest Classifier. It consists of many independent trees, or trees of choice in RF, that are trained with training sample data in the end. The results from each of these trees are then put through a voting process to produce estimates. An RF classifier is used to calculate the final result based on the majority of votes. Fig~\ref{fig:3} illustrates a block diagram of RF classifiers. We used GridSearchCV for hyperparamter tuning. The tuinted parameters are: \enquote{criterion}: \enquote{gini},\enquote{max\_depth}: 8,\enquote{max\_features}:\enquote{sqrt}, \enquote{n\_estimators}: 200. Along with that, five-fold cross-validation was applied to help in mitigating bias and variance issues and to ensure consistent model performance on unseen data.
The second model put into practice was the Decision Tree Classifier. The method is widely applied in machine learning to solve problems related to regression and classification. The internal node and the leaf node are the two types of nodes that a root node produces. Because they have several branches, internal nodes are known as decision-makers, whereas leaf nodes are known for producing results because they do not have any more branches. The fundamental design of the decision tree classifier is represented in Fig.~\ref{fig:4}. We used GridSearchCV for hyperparameter tuning. The tested parameters are: \enquote{criterion}: \enquote{entropy},\enquote{max\_depth}: 30,\enquote{min\_samples\_leaf}: 5,\enquote{min\_samples \_split}: 10,\enquote{max\_features}:\enquote{sqrt}. Along with that, five-fold cross-validation was applied to help in mitigating bias and variance issues and to ensure consistent model performance on unseen data.

\begin{figure}[h]
    \centering
    \includegraphics [width=0.45\textwidth] {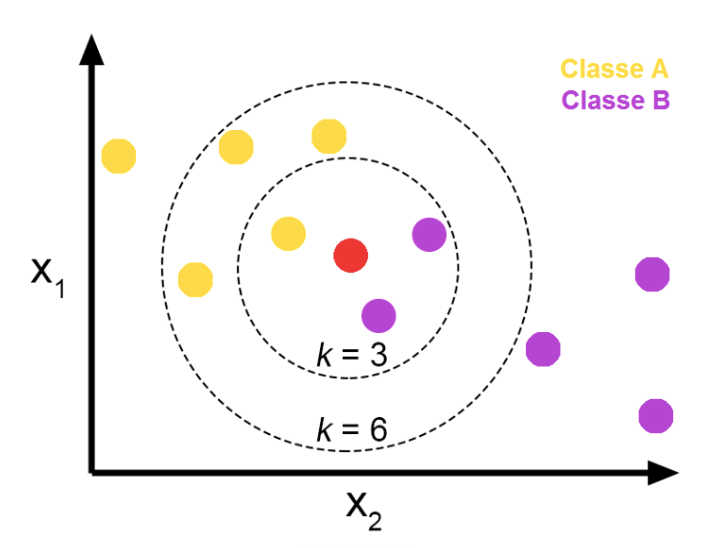}
    \caption{Diagram of K-Nearest Neighbor Classifier}
    \label{fig:5}
\end{figure}

\begin{figure}[h]
    \centering
    \includegraphics [width=0.45\textwidth] {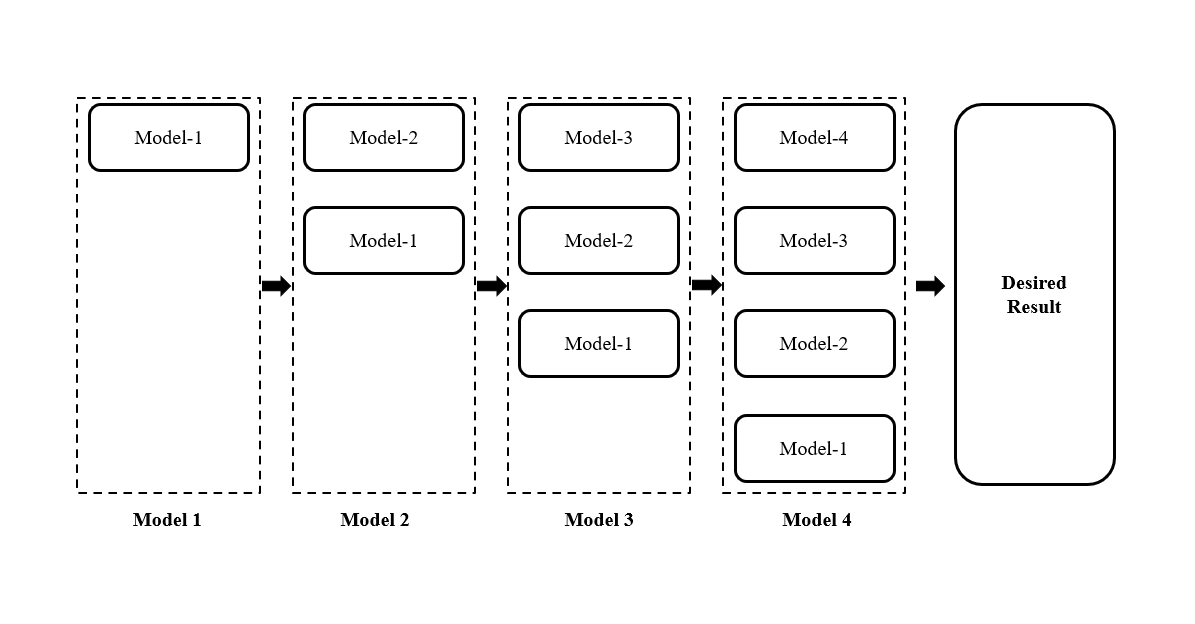}
    \caption{Diagram of Gradient Boosting Classifier}
    \label{fig:6}
\end{figure}

\begin{figure}[h]
    \centering
    \includegraphics [width=0.45\textwidth] {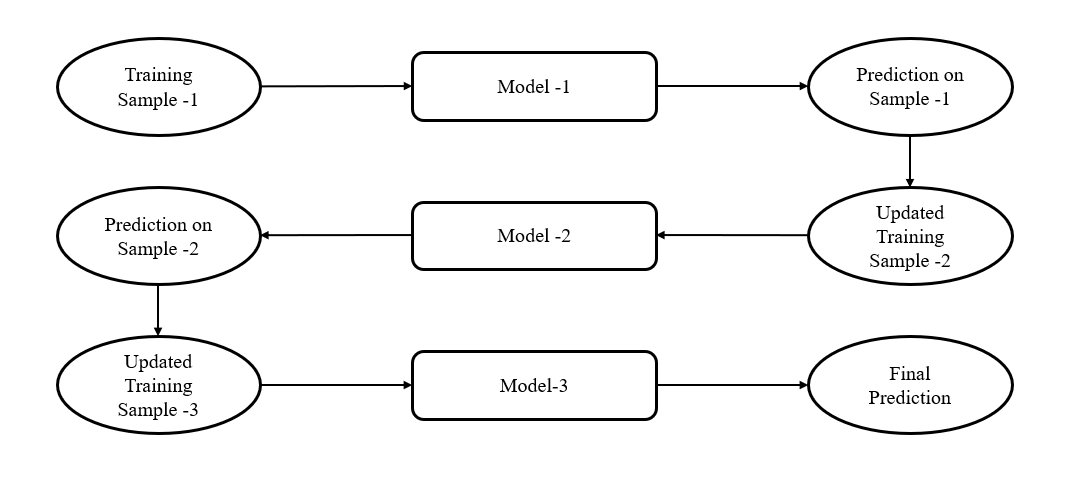}
    \caption{Diagram of Ada Boost Classifier}
    \label{fig:7}
\end{figure}

The K-Nearest Neighbor approach classifies new data points according to how similar they are to preexisting data points, saving all available data. This suggests that new data may be easily classified into an appropriate category by applying the KNN algorithm. In Fig.~\ref{fig:5}, the diagram of KNN is shown. We used GridSearchCV for hyperparamter tuning. The tuined parameters are: \enquote{n\_neighbors}: 5, \enquote{weights}: \enquote{distance},\enquote{metric}:\enquote{manhattan}, \enquote{p}: 1. Along with that, five-fold cross-validation was applied to help in mitigating bias and variance issues and to ensure consistent model performance on unseen data.
The gradient boosting algorithm (GBA) is one of the most effective machine learning techniques. Using this method, each predictor aims to minimize errors in order to outperform its prior performance. But the interesting idea behind gradient boosting is that it fits a new model to the regression that the previous predictor created instead of fitting a classifier to the data at each step~\cite{aliyev2020gradient}. Fig.~\ref{fig:6} shows the GB algorithm schematic design. We used GridSearchCV for hyperparamter tuning. The tested parameters are: \enquote{learning\_rate}: 0.01,\enquote{max\_depth}: 4, \enquote{n\_estimators}: 500, \enquote{subsample}: 0.8. Along with that, five-fold cross-validation was applied to help in mitigating bias and variance issues and to ensure consistent model performance on unseen data.
In machine learning, the AdaBoost algorithm is an ensemble method that uses boosting techniques. In this process, weights are re-allocated to each instance, with greater weights assigned to incorrectly identified instances. The construction of the first model and the method by which the algorithm finds errors in the first version are depicted in Fig.~\ref{fig:7}. We used GridSearchCV for hyperparameter tuning. The tuinted parameters are: \enquote{algorithm}: \enquote{SAMME.R}, \enquote{learning\_rate}: 0.1, \enquote{n\_estimators}: 100. Along with that, five-fold cross-validation was applied to help in mitigating bias and variance issues and to ensure consistent model performance on unseen data.

\subsection{Workflow Diagram}
\begin{figure}[h]
    \centering
    \includegraphics[width=0.45\textwidth]{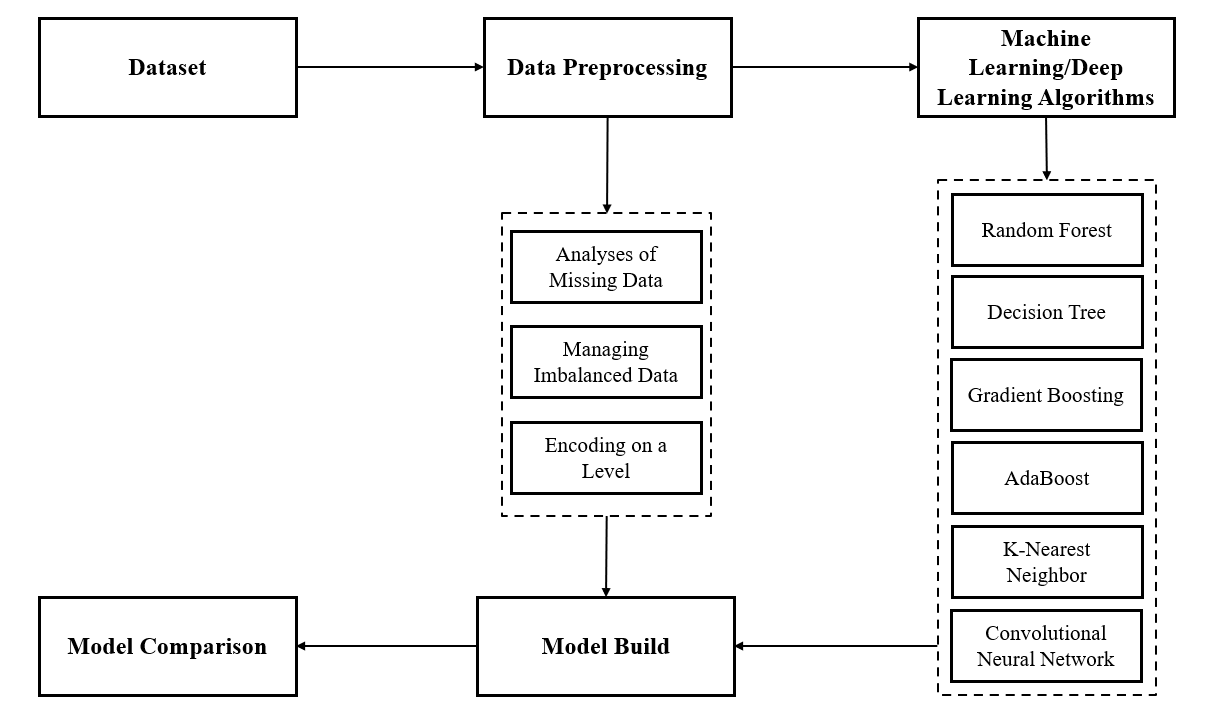}
    \caption {Workflow diagram of the system}
    \label{fig:2}
\end{figure}

Fig.~\ref{fig:2} represents the workflow diagram of this study. 
\subsection{Testing Methodology}
This subsection explains how the models were tested for this study. The equations of the testing metrics are given below~\cite{islam2024deep,mahmud2024advance}:
\begin{equation}
\label{eq:precision}
\text{Precision} = \frac{TP}{TP + FP}
\end{equation}

\begin{equation}
\label{eq:recall}
\text{Recall} = \frac{TP}{TP + FN}
\end{equation}

\begin{equation}
\label{eq:F1}
F1-score= 2 \times \frac{\text{Precision} \times \text{Recall}}{\text{Precision} + \text{Recall}}
\end{equation}

\begin{equation}
\label{eq:accuracy}
\text{Accuracy} = \frac{TP + TN}{TP + TN + FP + FN}
\end{equation}

\section{Result Analysis}
The models were evaluated based on precision, recall, f1 score, accuracy, confusion matrix, and ROC curves on the test set. The best two and the worst model's performance metrics, confusion matrix, and ROC curves are represented in this section.
\subsection{Random Forest}
\begin{figure}[h]
    \centering
    \includegraphics[width=0.45\textwidth]{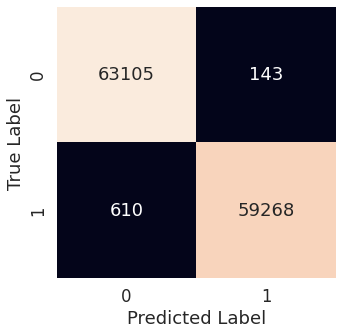}
    \caption{Confusion matrix of Random Forest Classifier}
    \label{fig:9}
\end{figure}

In Fig.~\ref{fig:9}, the confusion matrix of the random forest classifier is shown. The model's computed performance and the anticipated outcome are shown in the confusion matrix. The direct comparison of values such as true positives, false positives, true negatives, or false negatives is one advantage of using confusion matrices. This confusion matrix contains 59,268 true positives, 143 false negatives, 63,105 true negatives, 610 false negatives, 753 wrong predictions, and 122,373 right ones.

\begin{figure}[h]
    \centering
    \includegraphics[width=0.45\textwidth]{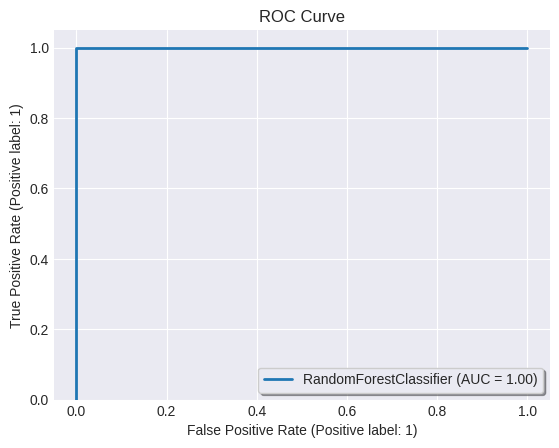}
    \caption{ROC curve of Random Forest Classifier}
    \label{fig:8}
\end{figure}

Fig.~\ref{fig:8} depicts the Random Forest classifier's ROC curve. With this model, the AUC value is 1.00. This indicates that the model makes very few incorrect predictions. The model performed almost accurately. 
\subsection{Decision Tree}
\begin{figure}[h]
    \centering
    \includegraphics[width=0.45\textwidth]{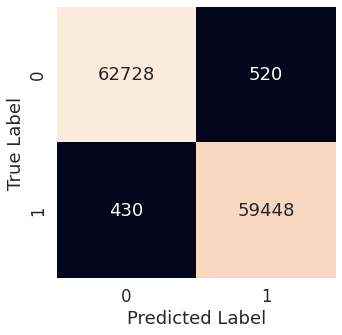}
    \caption{Confusion matrix of Decision Tree Classifier}
    \label{fig:11}
\end{figure}

\begin{figure}[h]
    \centering
    \includegraphics[width=0.45\textwidth]{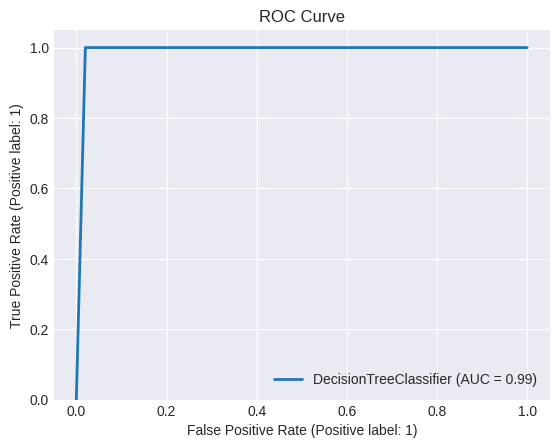}
    \caption{ROC curve of Decision Tree Classifier}
    \label{fig:10}
\end{figure}

Fig.~\ref{fig:11} shows the confusion matrix of the decision tree model. Using a confusion matrix, a classifier's predicted outcomes can be described. True positives, false positives, and true negatives can all be directly compared by using confusion matrices. This confusion matrix contains 59,448 true positives, 520 false negatives, 62,728 true negatives, and 430 false negatives. There were 122,176 correct and 950 incorrect predictions.
In Fig.~\ref{fig:10}, the ROC curve of the Decision Tree classifier is depicted. Here, the AUC value is 0.99. It indicates that the model made very few mistakes in terms of predicting attacks or normal situations. Unfortunately, it didn't perform as accurately as Random Forest.
\subsection{K-Nearest Neighbor}
\begin{figure}[h]
    \centering
    \includegraphics[width=0.45\textwidth]{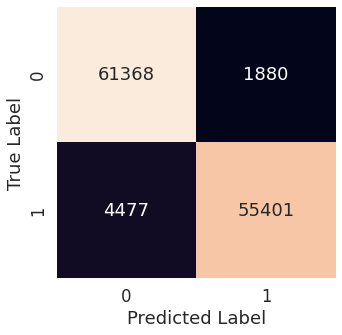}
    \caption{Confusion matrix of K-Nearest Neighbor classifier}
    \label{fig:13}
\end{figure}

Fig.~\ref{fig:13} depicts the K-Nearest Neighbor's confusion matrix. Using confusion matrices has the advantage of allowing for direct comparisons of variables like true positives, false positives, and false negatives, among others. There are 55,401 true positives, 1,880 false positives, 61,368 true negatives, and 4,477 false negatives in this confusion matrix, which indicates that 116,769 of the predictions were correct, while 6,357 were incorrect.

\subsection{Model Evaluation}
Table~\ref{tab:1} shows the precision, recall, and F1 score of the five applied machine learning models in this study. Clearly, in terms of all the performance metrics, Random Forest outperformed all the other models.

\begin{table}[htbp]
 \scriptsize
\caption{Performance Metrics Comparison}
\centering
\begin{tabular}{|c|c|c|c|}
\hline
\textbf{Model} & \textbf{Precision} & \textbf{Recall}&\textbf{F1 Score}\\
\hline
Random Forest & 0.997 & 0.989 & 0.994 \\ 
\hline
Decision Free & 0.991 & 0.993& 0.992 \\ 
\hline
Gradient Boosting & 0.982 & 0.974 & 0.978 \\ 
\hline
AdaBoost & 0.957 & 0.945 & 0.951 \\  
\hline
K-Nearest Neighbor &  0.967 & 0.925 & 0.945\\  
\hline
\end{tabular}
\label{tab:1}
\end{table}

Table~\ref{tab:RE} represents the performance results of the five applied machine learning models of our study. It is evident from the table that Random Forest outplayed all applied models in terms of accuracy and AUC. 

\begin{table}[htbp]
 \scriptsize
\caption{Result Evaluation of Models}
\centering
\begin{tabular}{|c|c|c|}
\hline
\textbf{Models} & \textbf{Accuracy} (\%) & \textbf{AUC}\\
\hline
Random Forest & \textbf{99.39} & \textbf{1.00} \\
\hline
Decision Free & 99.23 & 0.99 \\
\hline
Gradient Boosting & 97.89 & 0.98 \\
\hline
AdaBoost & 95.26 & 0.96 \\
\hline
K-Nearest Neighbor & 94.84 & 0.95\\
\hline
\end{tabular}
\label{tab:RE}
\end{table}

\subsection{Result Comparison}
As shown in Table~\ref{tab:2}, the models are compared to those previously studied. It is evident from the table that the Random Forest model overpowers all others in the framework.

\begin{table}[htbp]
\caption{Result Comparison}
\centering
\begin{tabular}{|c|c|c|c|c|}
\hline
\textbf{Study}  & \textbf{Best Model} & \textbf{Accuracy (\%)} \\
\hline
This paper & \textbf{Random Forest} & \textbf{99.39} \\
\hline
\cite{mohy2023ensemble} & Random Forest & 99.30\\
\hline
\cite{musleh2023intrusion} & VGG-16 combined with stacking & 98.3 \\
\hline
\cite{chaganti2023deep} & LSTM & 97.7 \\
\hline
\cite{khan2023hybrid} & Adamax & 97 \\
\hline
\end{tabular}
\label{tab:2}
\end{table}

\section{Conclusion and Future Work}
Right now, security is a significant concern for all organizations and establishments. In spite of the development of numerous methods to prevent intrusions into the network infrastructure, attempts are still being made by intruders to successfully access the data networks of these businesses and Web services. Intrusion detection systems (IDS) are developed to make it simpler for computer systems to deal with attacks. The development of a machine learning model may facilitate the earlier detection of intrusions and the prevention of severe consequences. This study focuses on how well various ML algorithms correctly anticipate intrusions based on a number of numerical variables. The Random Forest Classifier was able to give the highest accuracy of 99.39\% for this dataset. Although the model performed quite impressively, a limitation of this study remains. We were not able to perform this study on Bangladeshi IoT device attack/normal stage data. With a bigger dataset and machine learning models like ExtraTreesClassifier and Voting Classifier, the work may be enhanced in the future. Along with that, explainable AI (XAI) techniques can be added to do the model behavior analysis. As a result, the framework's consistency and demonstration will be improved. The intrusion detection systems of the future will interpret and apply these methods to detect new and emerging threats. In exchange for providing some fundamental information, the machine learning architecture can assist the general public in estimating the likelihood of a network intrusion. Real-world applications of this IDS span from securing smart homes and healthcare devices to safeguarding industrial control systems and automotive technologies, enhancing safety, and preventing unauthorized access and data theft.

\bibliographystyle{IEEEtran}
\bibliography{IEEE.bib}

\end{document}